\documentclass[a4paper]{jpconf}

\usepackage{graphicx}
\usepackage{amsmath}

\def\noeq#1{(\ref{#1})}
\def\1eq#1{Eq.~(\ref{#1})}

\def\2eqs#1#2{Eqs.~(\ref{#1}) and~(\ref{#2})}
\def\3eqs#1#2#3{Eqs.~(\ref{#1}),~(\ref{#2}) and~(\ref{#3})}

\def\fig#1{Fig.~\ref{#1}}

\def\s#1{{\scriptscriptstyle #1}}
\def\n#1{({\it #1}\,)}

\begin{document}
\title{The two-, three- and four-gluon sector of QCD in the Landau gauge}

\author{Daniele Binosi}

\address{European Centre for Theoretical Studies in Nuclear
Physics and Related Areas (ECT*) \\ and Fondazione Bruno Kessler,\\ Villa Tambosi, Strada delle
Tabarelle 286, 
I-38123 Villazzano (TN)  Italy}

\ead{binosi@ectstar.eu}

\begin{abstract}

Due to the nonperturbative masslessness of the ghost field, ghost loops that contribute to gluon Green's functions in the Landau gauge display infrared divergences, akin to those one would encounter in a conventional perturbative treatment. This is in sharp contrast with gluon loops, in which the perturbative divergences are tamed by the dynamical generation of a gluon mass acting as an effective infrared cutoff. In this paper, after reviewing the full nonperturbative origin of this divergence in the two-gluon sector, we discuss its implications for the three- and four-gluon sector, showing in particular that some of the form factors characterizing the corresponding Green's functions are bound to diverge in the infrared.        

\end{abstract}

\section{Introduction} In the past few years, the infrared (IR) behavior of Yang-Mills Green's functions in the Landau gauge has been the subject of numerous studies  both in the continuum and on the lattice. Thanks to the use of a variety of different theoretical approaches (see, {\it e.g.},~\cite{Binosi:2009qm} and references therein) as well as extensive numerical simulations on large lattices~\cite{Cucchieri:2007md,Cucchieri:2010xr,Bogolubsky:2009dc,Ayala:2012pb}, a consistent picture has unequivocally emerged for the different $n$-point sectors of the theory.

Most notably, it has been firmly established that the gluon propagator saturates at small momenta in a way consistent with the presence of a dynamically generated gluon mass~\cite{Cornwall:1981zr,Aguilar:2011ux,Binosi:2012sj,Aguilar:2014tka}; the ghost propagator is instead essentially free in the same momentum region: in this case it is the ghost {\it dressing function} (defined as $q^2$ times the propagator, see below) that saturates to a finite non-vanishing value~\cite{Boucaud:2008ky,Aguilar:2008xm}.

This characteristic behavior of the two-point sector, which has been found to be valid in three and four space-time dimensions, for SU(3) and SU(2) gauge groups, and with or without the inclusion of dynamical quarks, profoundly affects the IR behavior of the theory's {\it entire} tower of $n$-point Green's functions~\cite{Aguilar:2013vaa}. In fact, it turns out that $n$-point functions exhibiting an `ancestor' ghost-loop ({\it i.e.},  a ghost-loop that is originally present in the lowest order perturbative expansion of the function under scrutiny), will develop a logarithmic IR singularity: the contribution of such diagram will correspond in fact to a pure logarithm, $\log q^2/\mu^2$ (where $\mu$ is some chosen scale), which is unprotected, in the sense that there is no mass term in its argument that could tame the corresponding divergence in the low momenta region. On the contrary, if the lowest order loops have a circulating gluon (and therefore a gluon propagator appears), the corresponding logarithm will be of the type $\log (q^2-m^2)/\mu^2$; thus, due to the presence of the dynamically generated gluon mass, such contributions will be finite for arbitrary low momenta. 

In the following, after reviewing the full nonperturbative analysis of the divergent ancestor ghost loop appearing in the 2-point sector and its effects on the gluon propagator and (inverse) dressing function, we will discuss the implications for the gluon three- and four-point sectors. In particular, we will show that the presence of these loops implies the appearance of IR divergences in some of the form factors comprising the general Lorentz decomposition of the corresponding  vertices~\cite{Aguilar:2013vaa,Binosi:2014kka}.

\section{\label{sect:2-point}The 2-gluon sector}

The so-called PT-BFM framework~\cite{Aguilar:2006gr,Binosi:2007pi,Binosi:2008qk}, originating from the combination of  the Pinch Technique (PT)~\cite{Cornwall:1981zr,Cornwall:1989gv,Binosi:2002ft,Binosi:2003rr} 
	with the Background Field Method (BFM)~\cite{Abbott:1981ke}, turns out to be particularly suited for studying the problem at hand, as it allows to separate in a gauge invariant way the ghost and gluon contributions to Green's functions. Consequently, one is able to isolate ancestor ghost loops and study their IR behavior in a meaningful way.  

\begin{figure}[!t]
\includegraphics[scale=1.1]{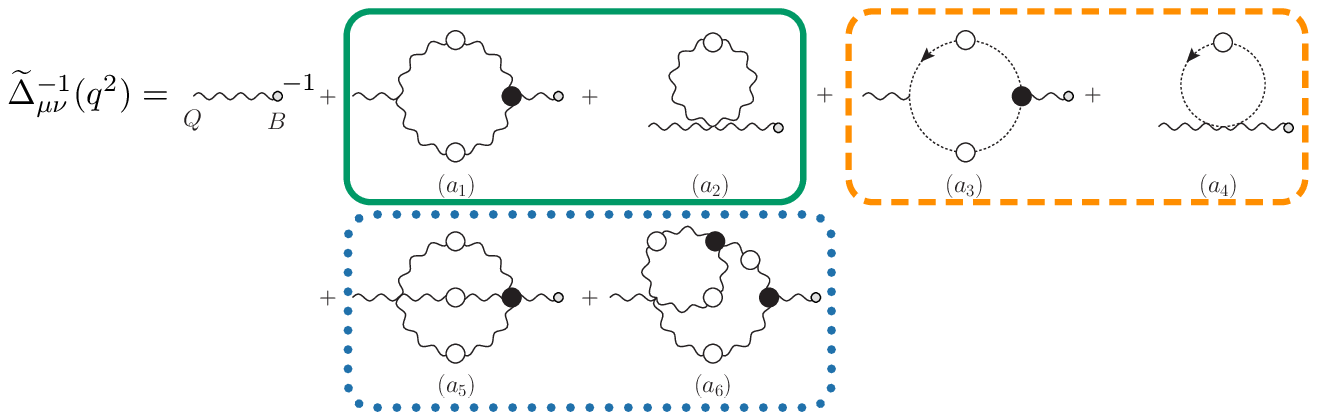} 
\caption{\label{fig:QB-SDE}(color online). The SDE obeyed by the $QB$ gluon propagator. 
Each of the three different boxes (continuous, dashed, and dotted line) contains an individually transverse subgroup of diagrams, 
{\it i.e.}, $q^{\mu} [(a_i) + (a_{i+1})]_{\mu\nu} = 0$ with $i=1,2,3$. 
Black (white) blobs represent fully dressed 1-PI (connected) Green's functions; finally, small gray circles appearing on the external legs indicate  background gluons.}
\end{figure}

Within the PT-BFM framework, one  considers the Schwinger-Dyson equation (SDE) of~\fig{fig:QB-SDE} describing the propagator $\widetilde{\Delta}_{\mu\nu}(q)$ of a quantum ($Q$) and a background ($B$) gluon. Writing for the conventional ({\it i.e.}, $QQ$) propagator
\begin{equation}
	i\Delta_{\mu\nu}(q)=-iP_{\mu\nu}(q)\Delta(q^2);\qquad P_{\mu\nu}(q)=g_{\mu\nu}-q_\mu q_\nu/q^2,
\end{equation}  
and similarly for $\widetilde{\Delta}_{\mu\nu}(q)$, it turns out that the two propagators are related by a so-called Background-Quantum identity (BQI), reading (see again~\fig{fig:QB-SDE})
\begin{equation}
	[1+G(q^2)]\Delta^{-1}(q^2){P}_{\mu\nu}(q) = \widetilde{\Delta}^{-1}(q^2)P_{\mu\nu}(q)=
q^2 {P}_{\mu\nu}(q) + i\,\sum_{i=1}^{6}(a_i)_{\mu\nu}.
\label{2p-bqi}
\end{equation}
The auxiliary function $G(q^2)$ appearing above, corresponds to the metric form factor of a special Green's function that is typical of this framework and describes the ghost-gluon dynamics. If one introduces the ghost dressing function $F(q^2)$ through
\begin{equation}
	F(q^2)=q^2D(q^2),
\end{equation}
 $D(q^2)$ being the full ghost propagator, one has the approximate identity~\cite{Grassi:2004yq,Aguilar:2009nf}
\begin{equation}
	1+G(q^2)\approx F^{-1}(q^2),
	\label{app}
\end{equation}
a relation that becomes exact at $q^2=0$.

The advantage of employing the BQI~\noeq{2p-bqi}, and therefore considering the $BQ$ self-energy diagrams rather than the $QQ$ ones, resides in the fact that owing to the background Ward identity, all subsets of graphs enclosed within each box of~\fig{fig:QB-SDE} give rise to a transverse contribution~\cite{Aguilar:2006gr,Binosi:2007pi,Binosi:2008qk}. Thus, their individual treatment, or, in fact, the total omission of  entire subsets from one's analysis, does not tamper with the transversality of the gluon self-energy\footnote{In fact, it has been shown in~\cite{Binosi:2013cea} that the PT-BFM framework is not an {\it ad-hoc} procedure, rather it naturally emerges from the requirement of antiBRST invariance of the SU(N) Yang-Mills action. The additional identities (background Ward identity, BQIs, {\it etc.}) present in this novel formulation, are then none but manifestations of the underlying BRST-antiBRST invariance of the theory.}.  

Now, in the presence, of an IR finite gluon propagator, the function $\Delta(q^2)$ can be decomposed as
\begin{equation}
	\Delta^{-1}(q^2)=q^2J(q^2)+m^2(q^2),
\end{equation}
where $J(q^2)$ is the inverse of the gluon dressing function, while $m^2(q^2)$ is the dynamically generated gluon mass. Notice that~\1eq{2p-bqi} is satisfied independently by $J(q^2)$ and $m^2(q^2)$, so that one has
\begin{equation}
	X(q^2)=F(q^2)\widetilde{X}(q^2);\qquad X(q^2)=J(q^2), m^2(q^2),
	\label{ind-BQI}
\end{equation}
where we have used the approximation~\noeq{app}.
 
The block-wise transversality property of the $BQ$ propagator together with the BQI~\noeq{ind-BQI},  allows then for a meaningful separation of two kinds of contribution to $J(q^2)$: the ones stemming from the ghost graphs $(a_3)$ and $(a_4)$, and the ones stemming from the remaining gluon graphs. Denoting them respectively by $J_c(q^2)$ and $J_g(q^2)$, 
	one has therefore
\begin{eqnarray}
	q^2J_g(q^2)P_{\mu\nu}(q)&=&F(q^2)[(a_1)+(a_2)]_{\mu\nu}+F(q^2)[(a_5)+(a_6)]_{\mu\nu},,\nonumber \\
	q^2J_c(q^2)P_{\mu\nu}(q)&=&F(q^2)[(a_3)+(a_4)]_{\mu\nu}
	\end{eqnarray}  
where on the right-hand side we assume that one is evaluating only the terms that vanishes as $q^2$ goes to zero (the non vanishing terms contributing instead to the mass equation, see~\cite{Aguilar:2011ux,Binosi:2012sj}). Then one finally has
\begin{equation}
	J(q^2)=1+J_g(q^2)+J_c(q^2),
	\label{J}
\end{equation}
with the ``1'' corresponding to the tree-level term.

It turns out that there is a profound difference in the behavior of the gluon inverse dressing functions $J_c(q^2)$ and $J_g(q^2)$, which ultimately reflects the fact that the particle circulating in the loops of the corresponding diagrams have completely different behavior in the IR. In fact, in this momentum regime ghosts behave like massless free particles, $D(q^2)\sim F(0)/q^2$, whereas gluons are effectively massive, and therefore $\Delta(q^2)\sim1/m^2(0)$. 

This difference between the inverse dressings $J_c(q^2)$ and $J_g(q^2)$ is most readily understood at the lowest order in the perturbative expansion: in the four-dimensional case, $J_g(q^2)$ develops a logarithm tamed by the presence  of the effective IR cutoff provided by the dynamical gluon mass; on the contrary, $J_c(q^2)$ displays an unprotected logarithm, which vanishes at a finite value of $q^2$, then reverses its sign, becoming finally divergent at the origin\footnote{In the three-dimensional case the divergence in $J_c(q^2)$ is linear in $q$ with $J_c(q^2)\sim1/q$ while for the gluon $J_g(q^2)\sim\arctan q/2m$.}. As a consequence, one can easily show that the gluon propagator must display a maximum (and, consequently, its inverse a minimum) located in the deep IR region~\cite{Aguilar:2013vaa}.
     
Consider now the full non perturbative case. The fact that the background gluon-ghost vertex appearing in the ghost block of~\fig{fig:QB-SDE} satisfies a QED-like Ward identity, furnishes a closed all-order expression for the longitudinal part of this vertex (which is not possible to obtain for the conventional gluon-ghost vertex). This leaves the transverse (automatically conserved) part of the vertex undetermined; however, under mild assumptions on the behavior of the form factor characterizing the latter vertex structure, one can show that the neglected term would give rise only to IR subleading contributions to $J_c(q^2)$ (see~\cite{Aguilar:2013vaa} for details).

Following the analysis presented in~\cite{Aguilar:2013vaa}, one then finds that the ghost diagrams contribute to the gluon inverse dressing function the term
\begin{equation}
	q^2J_c(q^2)= \frac{g^2N}{2(d-1)} \, F(q^2)\left[4T(q^2)+q^2S(q^2)\right],
\end{equation}
where
\begin{eqnarray}
	T(q^2)&=&
\int_k \frac{F(k+q)-F(k)}{(k+q)^2-k^2}+\left(\frac{d}2-1\right)\int_k\frac{F(k)}{k^2},\nonumber \\
S(q^2)&=&\int_k\frac{F(k)}{k^2(k+q)^2}-\int_k\!\frac{F(k+q)-F(k)}{k^2[(k+q)^2-k^2]}. 
\label{SandT-final}
\end{eqnarray}
In the equations above $N$ represents the number of colors, and we have introduced the $d$-dimensional measure $\int_{k}\equiv \mu^{\epsilon}/(2\pi)^{d}\!\int\!\mathrm{d}^d k$, with $\mu$  the 't Hooft mass and $\epsilon=4-d$. Then, using the fact that $T(0)=0$, in the deep IR region one finds the following behavior~\cite{Aguilar:2013vaa} 
\begin{eqnarray}
	T(q^2) &\underset{q^2\to0}{\to}&\ -\frac{1}{12}\left(d-2\right)q^2\!\int_k \frac{1}{k^2}\frac{\partial F(k)}{\partial k^2} + {\cal O}(q^4),\nonumber \\
	S(q^2) &\underset{q^2\to0}{\to}&\int_k \frac{F(k)}{k^4} - \int_k \frac{1}{k^2}\frac{\partial F(k)}{\partial k^2} + {\cal O}(q^2).
\end{eqnarray}
The first integral appearing in the expansion of $S(q^2)$, contains the expected logarithmic divergence, as it can be easily seen by setting  $F(k^2) = 1$.  Since the full $F(k^2)$ saturates at a constant value in the IR, its presence will not qualitatively modify the behavior of the integral; it will rather change its prefactor from 1 to $F(0)$.  The second integral in $S(q^2)$, and therefore $T(q^2)$, is subleading:  introducing spherical coordinates and integrating by parts, makes manifest that this integral is simply proportional to $F(0)$. In fact, \1eq{J} can be rewritten as
\begin{equation}
	J(q^2) = J_c^\ell(q^2) + J^{\mathrm{s}\ell}(q^2),
\end{equation}
where
\begin{equation}
	J_c^\ell(q^2)=\frac{g^2N}{2(d-1)} F(q^2)\int_k\frac{F(k)}{k^2(k+q)^2},
	\label{Jclead}
\end{equation}
whereas $J^{\mathrm{s}\ell}(q^2)$ represents the IR subleading terms (including the terms generated by gluon graphs).

\begin{figure}[!t]
\mbox{}\hspace{-1.4cm}\includegraphics[scale=.975]{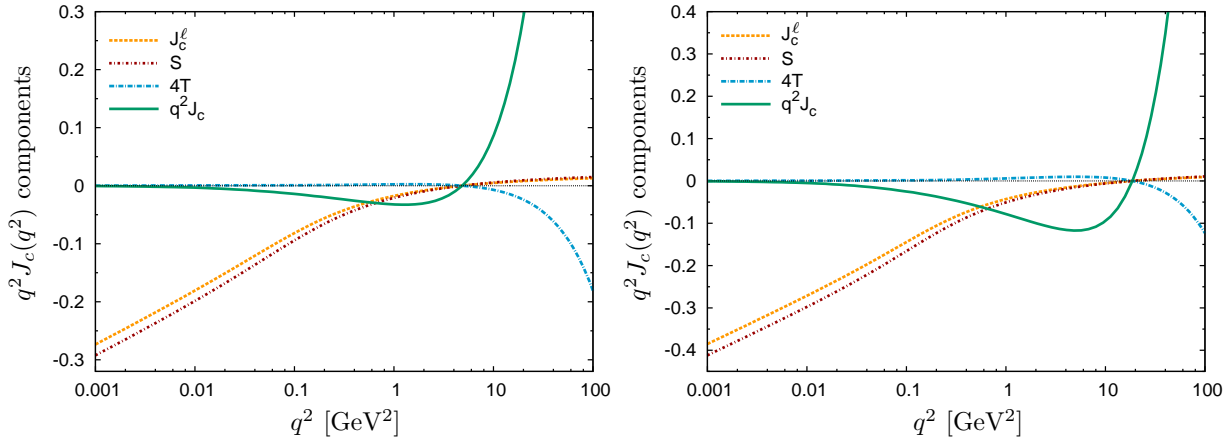}
\caption{\label{fig:q2Jc4d-4d}(color online). The ghost-loop contribution,  $q^2J_c(q^2)$, to the full gluon kinetic term $q^2J(q^2)$ for the SU(2) (left) and SU(3) (right) gauge groups.}
\end{figure}

The terms appearing in~\1eq{SandT-final} can be next evaluated numerically by using as input a functional fit to the SU(2)~\cite{Cucchieri:2010xr} and SU(3)~\cite{Bogolubsky:2009dc} quenched lattice data for the ghost dressing function. The results are shown in~\fig{fig:q2Jc4d-4d}, where, as anticipated, the IR logarithmic divergence is clearly identified by the linear behavior (in log scale) of the $S$ term above.   

To check whether or not the nonperturbative propagator has a maximum as a consequence of the divergence of its ghost contribution $J_c^\ell(q^2)$~\noeq{Jclead}, let us consider the derivative of the inverse gluon propagator that reads
\begin{equation}
	[\Delta^{-1}(q^2)]'=[q^2J(q^2)]'+m'(q^2)=J_c^\ell(q^2) +  
[ J^{\mathrm{s}\ell}(q^2)+ q^2 J^{\,\prime}(q^2)]-m'(q^2).
\end{equation} 

Evidently, the quantity in brackets is subleading in the IR, while the fact that the propagator is decreasing in the UV, ensures that the above derivative is positive in this region. In addition, the dynamical equation governing  $m^2(q^2)$ is known~\cite{Aguilar:2011ux,Binosi:2012sj,Aguilar:2014tka} and its solutions are monotonically decreasing and possess a finite derivative in the origin. Thus we conclude that the derivative above must reverse the sign at a point $q_\s\Delta$ where the propagator will display a maximum. 

\begin{figure}[!t]
\mbox{}\hspace{-0.5cm}
\includegraphics[scale=.975]{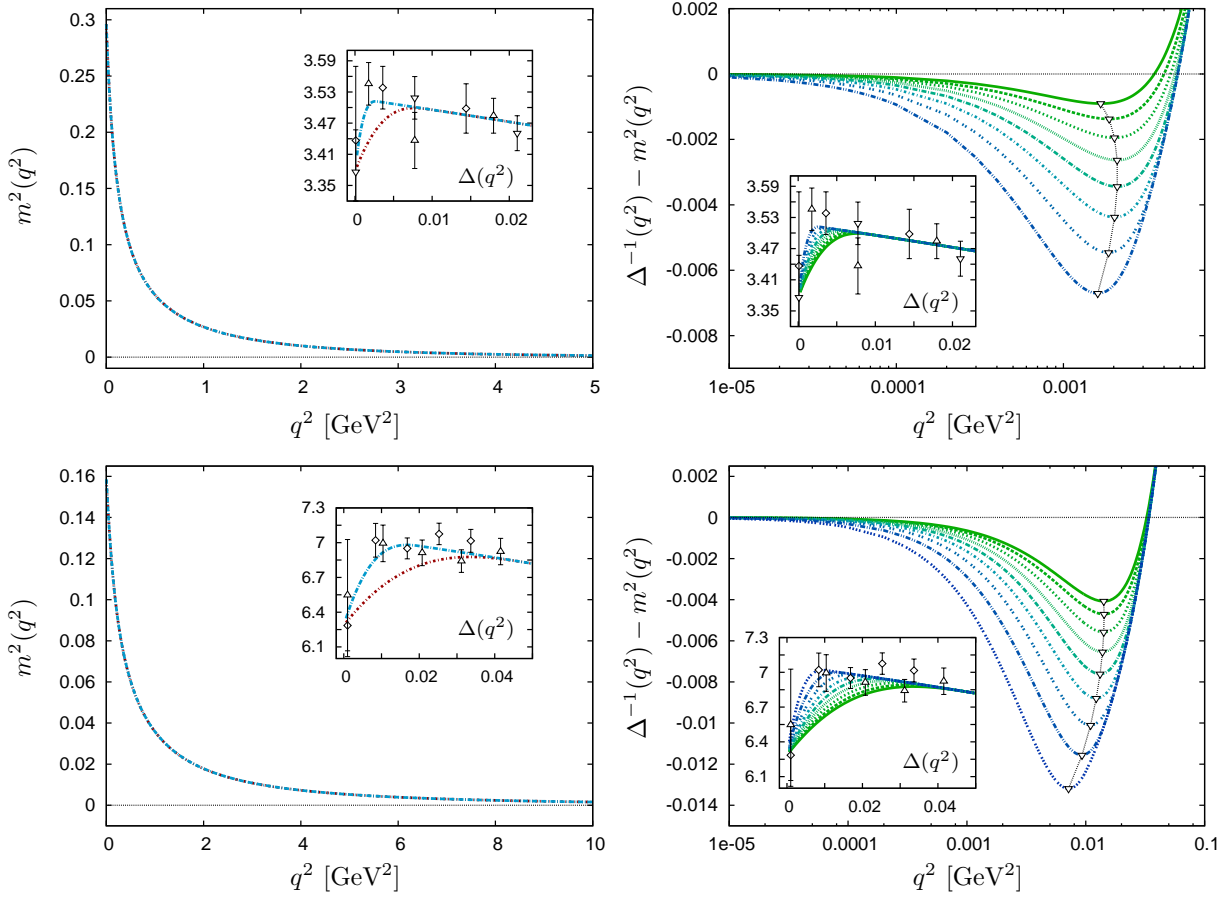}
\caption{\label{fig:mass-lake-4d}(color online). The dynamical gluon mass (left panels) and the propagator's full kinetic part $\Delta^{-1}(q^2)-m^2(q^2)$ (right panels) for the SU(2) (top) and SU(3) (bottom) gauge groups. Whereas the solutions of the mass equation are clearly insensitive to the presence of a maximum in the propagator, as shown here for two representative cases, the full kinetic term develops a negative minimum ($q_\s{J}$), whose position is marked in the right panels by open (down) triangles. Insets show in all cases the IR behavior of the various propagator fits used as input, together with the corresponding lattice data of Refs.~\cite{Cucchieri:2007md,Cucchieri:2010xr} and~\cite{Bogolubsky:2009dc}.} 
\end{figure}

Thus, according to the reasoning developed so far, and as a consequence of the masslessness of the ghost field, the lattice data for the gluon propagator ought to display a maximum, located in the (deep) IR region. As displayed in the insets appearing on the right panels of~\fig{fig:mass-lake-4d}, such a maximum appears to be indeed encoded in the lattice data for $\Delta$, which reveals a suppression of the deep IR points independently from the gauge group chosen.  In those same insets we also plot different fitting curves in which the position of the maximum is varied.

For reasons that will become clear in the next section, it is interesting to study the full kinetic term $q^2J(q^2)$, an indirect knowledge of which\footnote{The reason why we do not perform a direct SDE analysis of this quantity  is because we do not have a satisfactory control over some of the basic ingredients appearing in the integral equation governing  $J(q^2)$; in particular, and most notably contrary to what happens for the Landau gauge mass equation, the equation for $J(q^2)$ involves the fully-dressed four-gluon vertex, whose structure is presently poorly known (see also Sect.~\ref{sect:4-point}).} can be acquired by evaluating the combination
\begin{equation}
	q^2J(q^2)=\Delta^{-1}(q^2)-m(q^2),
	\label{ind-kinetic}
\end{equation}  
where $\Delta(q^2)$ is obtained from the aforementioned fits to the lattice while $m^2(q^2)$ is obtained by solving the associated mass equation. Notice that one expects that the quantity~\noeq{ind-kinetic} develops a minimum at a location $q_\s{J}$, which in general however will not coincide with $q_\s\Delta$. 

The indirect determination of $q^2 J(q^2)$ from \1eq{ind-kinetic}, using as basic input  the family of curves for $\Delta(q^2)$ obtained in the previous step is shown in~\fig{fig:mass-lake-4d}. First we established that, when the latter curves are used as input to determine the solution to the  mass equation, the resulting masses are completely independent of the location and the size of the maximum of the propagator (left panels of~\fig{fig:mass-lake-4d}). Once the combination~\noeq{ind-kinetic} is formed (right panels of~\fig{fig:mass-lake-4d}), we observe that the full kinetic term obtained vanishes at the origin, decreases in the deep IR, and reaches a negative minimum before crossing zero and turning positive (we mark for each curve the location of the corresponding minimum, $q_{\s J}$). 

Summarising, the fact that the ghost field remains  nonperturbatively massless, as opposed to the  gluon which acquires  a  dynamically   generated  mass,  implies  unavoidably  the existence of a negative IR divergence in the dimensionless co-factor  $J(q^2)$ of the  kinetic part of the gluon propagator. While this divergence, which originates exclusively from one-loop dressed diagrams involving a ghost loop,  does not spoil the overall  finiteness of the gluon two-point function, it affects it in two different ways: \n{i}  it forces the appearance of a maximum in the gluon propagator located at $q_\s{\Delta}$, and, correspondingly, \n{ii} it makes the full kinetic term develop a minimum at a location $q_\s{J}$.  

\section{The $n$-gluon sector}

\begin{figure}[!t]
\centerline{\includegraphics[scale=.64]{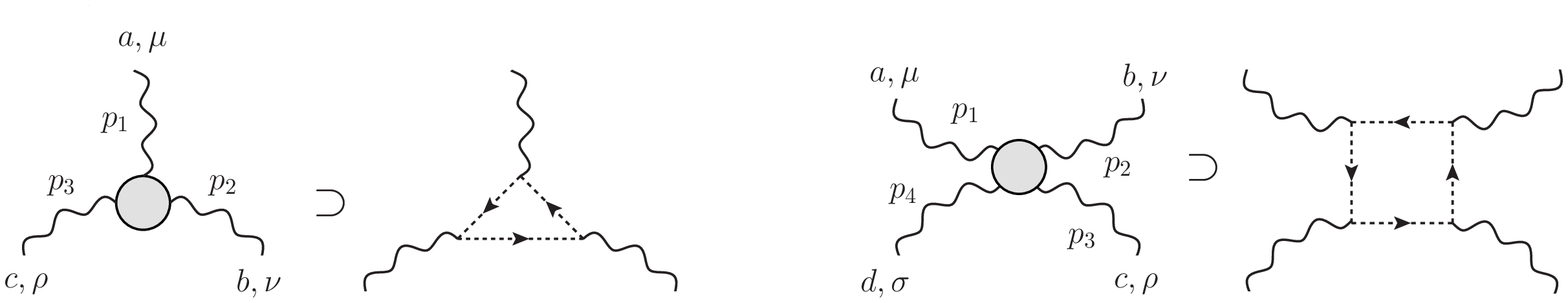}}
\caption{\label{fig:3-and-4-gluon}Ancestor ghost loops in the case of the three- and four-gluon sector. For later convenience we show the color and Lorentz indices of the external legs together with the momentum flow (all momenta are entering).}
\end{figure}

The generality of the analysis performed in the previous section, suggests  that the IR divergence appearing in the two-point sector, is likely to manifest itself in other Green's  functions  that contain  an ancestor ghost  loop. Natural candidates are clearly the three- and four-point functions, due to the presence respectively of the triangular- and box-like diagrams shown in~\fig{fig:3-and-4-gluon}.

In order to study this issue, in what follows we resort to the quantity usually employed at the non-perturbative level for studying $n$-point functions in the Landau gauge, namely the ratio
\begin{equation}
	R_n^\s{T}(q_1,\cdots,q_n)=\frac{T^{a_1\cdots a_n}_{\mu_1\cdots\mu_n}(q_1,\cdots,q_n) P^{\mu_1\nu_1}(q_1)\cdots P^{\mu_n\nu_n}(q_1)\Gamma^{a_1\cdots a_n}_{\nu_1\cdots\nu_n}(q_1,\cdots,q_n)}{T^{a_1\cdots a_n}_{\mu_1\cdots\mu_n}(q_1,\cdots,q_n) P^{\mu_1\nu_1}(q_1)\cdots P^{\mu_n\nu_n}(q_1)T^{a_1\cdots a_n}_{\nu_1\cdots\nu_n}(q_1,\cdots,q_n)}.
	\label{genral-R}
\end{equation}
What~\1eq{genral-R} achieves is therefore the projection of the full vertex $\Gamma^{a_1\cdots a_n}_{\nu_1\cdots\nu_n}(q_1,\cdots,q_n)$ under scrutiny onto a particular tensor structure $T^{a_1\cdots a_n}_{\mu_1\cdots\mu_n}(q_1,\cdots,q_n)$, factoring out at the same time external leg corrections\footnote{This definition allows, when data are available, for a direct comparison with lattice results, as in this case only connected (as opposed to 1-PI) Green's functions can be measured.}. Such a ratio depends generally by the modulo of the $n-1$ independent momenta and the $(n-1)(n-2)/2$ angles between them.

\subsection{\label{sect:3-point}The three-gluon sector}

We start by considering the case of the three gluon vertex, defined according to (all momenta entering)
\begin{eqnarray}
	\Gamma_{A^a_\mu A^b_\nu A^c_\rho}(p_1,p_2,p_3)&=&-ig\Gamma^{abc}_{\mu\nu\rho}(p_1,p_2,p_3);\nonumber \\
	\Gamma^{(0)abc}_{\mu\nu\rho}(p_1,p_2,p_3)&=&f^{abc}[g_{\mu\nu}(p_1-p_2)_\rho+g_{\nu\rho}(p_2-p_3)_\mu+g_{\rho\mu}(p_3-p_1)_\nu],
\end{eqnarray}
where $f^{abc}$ are the real and totally antisymmetric structure constants, satisfying the normalization condition $f^{amn}f^{bmn}=N\delta^{ab}$.

Choosing to project the vertex on its tree-level tensor structure, and considering for the momenta the so-called orthogonal configuration, corresponding to setting the angle between $p_1$ and $p_2$ to $\pi/2$ and then taking the limit $p_2^2\to0$, it can be shown that~\cite{Aguilar:2013vaa}
\begin{equation}
	R_3^{\Gamma^{(0)}}[q^2,0,\pi/2]=F(0)[q^2J(q^2)]'+R^{\mathrm{s}\ell}(q^2),
	\label{R3}
\end{equation} 
where we have set $q^2=p_1^2$, and the last term collects all subleading corrections not contained in the first one. Then the IR behavior of the ratio~\noeq{R3} is driven once again by the gluon inverse dressing function so that
\begin{equation}
	R_3^{\Gamma^{(0)}}[q^2,0,\pi/2]\underset{q^2\to0}{\sim}F(0)J_c^\ell(q^2).
	\label{R-J}
\end{equation}
Thus, the expectation is that  $R_3^{\Gamma^{(0)}}$ in the orthogonal configuration will vanish at a point $q_0$ an then have a negative logarithmic IR divergence\footnote{For related studies on the three-gluon vertex, see~\cite{Pelaez:2013cpa,Blum:2014gna,Eichmann:2014xya}.}. Notice that an estimate for $q_0$ is provided by $q_\s{J}$ as the relation~\noeq{R3} reveals; in particular, for the SU(2) gauge group the minimum of the full kinetic term provides the estimate $q_0\approx45$ MeV, while for SU(3) we obtain $q_0\approx130$ MeV. 

\begin{figure}[!t]
\includegraphics[scale=0.975]{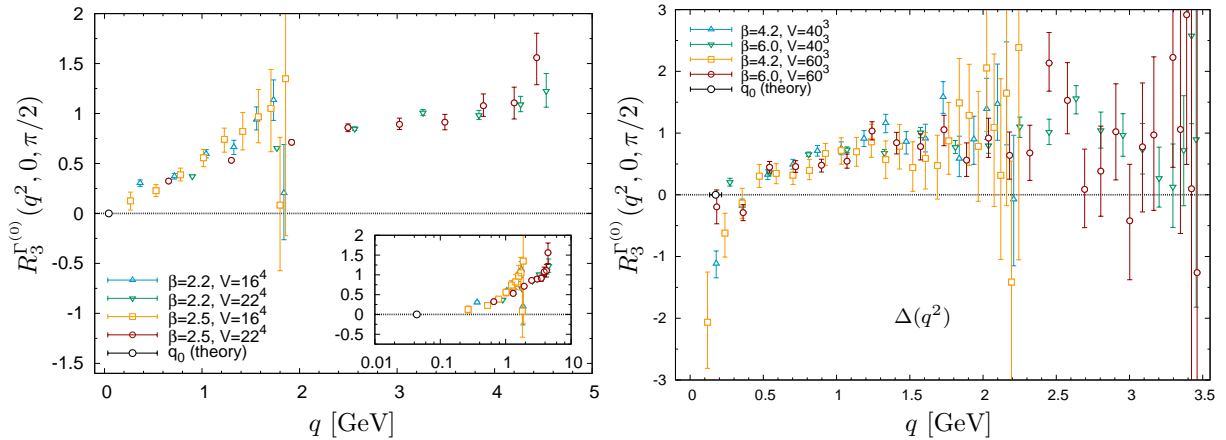} 
\caption{\label{fig:3gluon-3d-4d}(color online). {\it Left panel:} Prediction for the zero-crossing compared with the SU(2) ratio $R_3^{\Gamma^{(0)}}(q^2,0,\pi/2)$ measured on the lattice in four-dimensions~\cite{Cucchieri:2008qm}. The inset shows a logarithmic plot of the same quantity. {\it Right panel:} Same thing, but for the three-dimensional case~\cite{Cucchieri:2008qm}; notice that in this case the ghost divergence is linear in $q$, and therefore the lattice is able to resolve both the zero crossing and the negative divergence.}
\end{figure}

Our SU(2) result is compared with  the behavior of $R_3^{\Gamma^{(0)}}(q^2,0,\pi/2)$ obtained from lattice simulations~\cite{Cucchieri:2008qm} on the left panel of \fig{fig:3gluon-3d-4d}; as one can see while there is indeed an indication that the zero crossing is going to happen, the actual value is located too deep in the IR to be resolved with current lattice volumes. On the other hand, we show on the right panel of the same figure the three-dimensional case, where the leading ghost divergence $J_c^\ell(q^2)$ is linear in momentum rather than logarithmic: in this case the zero crossing and divergent behavior are clearly resolved by the lattice and our prediction $q_0\approx175$ MeV compares reasonably well with the available  data~\cite{Cucchieri:2008qm}.

\subsection{\label{sect:4-point}The four-gluon sector}

We next turn our attention to the four-gluon vertex which is defined according to (all momenta entering)
\begin{eqnarray}
\Gamma_{A^a_\mu A^b_\nu A^c_\rho A^d_\sigma}(p_1,p_2,p_3,p_4)&=&-ig^2\Gamma_{\mu\nu\rho\sigma}^{abcd}(p_1,p_2,p_3,p_4);
\nonumber \\
\Gamma_{\mu\nu\rho\sigma}^{abcd(0)}&=&f^{adr}f^{cbr}(g_{\mu\rho}g_{\nu\sigma}-g_{\mu\nu}g_{\rho\sigma})+f^{abr}f^{rdc}(g_{\mu\sigma}g_{\nu\rho}-g_{\mu\rho}g_{\nu\sigma})\nonumber \\
&+&f^{acr}f^{dbr}(g_{\mu\sigma}g_{\nu\rho}-g_{\mu\nu}g_{\rho\sigma}).
\label{4g}
\end{eqnarray}
As already remarked this is the most poorly understood vertex of the theory, {\it e.g.}, no lattice simulation of this quantity exists to date (and, consequently, no data on any ratio $R_n^\s{T}$ in any momentum configuration are available). 

\begin{figure}[!t]
\centerline{\includegraphics[scale=0.8]{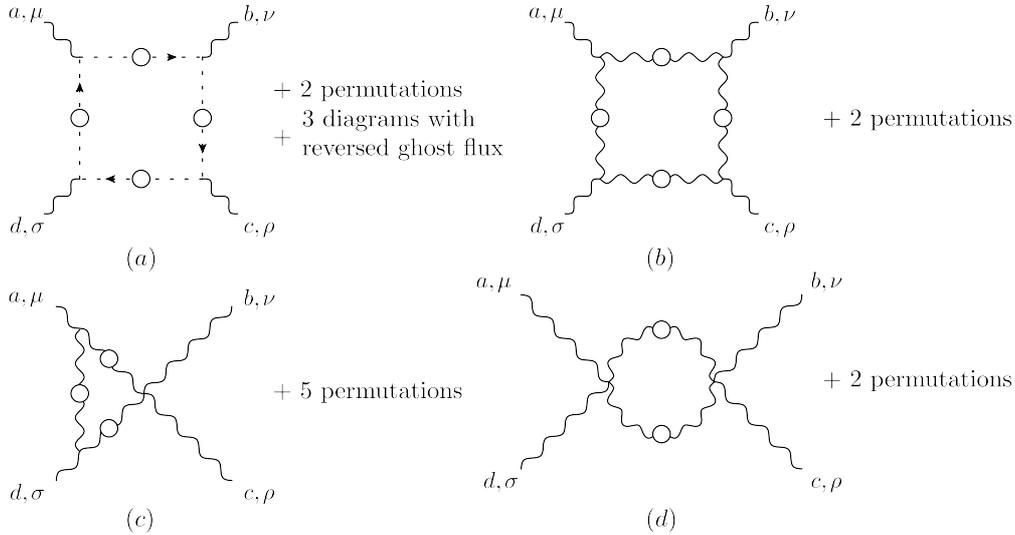}}\caption{\label{fig:4g-1loop-dressed}The 18 diagrams contributing to the four-gluon vertex in the one-loop dressed approximation. The diagrams are divided in four different classes: $(a)$ ghost boxes, $(b)$ gluon boxes, $(c)$ triangles, $(d)$ fishnets (which carry a statistical factor of $1/2$).}
\end{figure}

However, motivated by our successful description of the 2- and three-gluon sector, a preliminary nonperturbative study of this vertex can be attempted~\cite{Binosi:2014kka}. To this purpose, one can resort to a somewhat simplified methodology, in which the different form factors are extracted directly from the evaluation of one-loop diagrams with fully dressed propagators but tree-level vertices (\fig{fig:4g-1loop-dressed}). 

Even within this simplified setting, the calculation of the 18 one-loop dressed diagrams of~\fig{fig:4g-1loop-dressed} in a general momentum configuration would be a challenging task, due to the vast proliferation of tensorial structures. Indeed, at the level of rank-4 Minkowski tensors one has schematically the structures $gg$, $gp_ip_j$ and $p_ip_jp_kp_m$, whereas for the rank-4 color tensors the possible combinations are of the type $f\!f$, $dd$, $f\!d$ and $\delta\delta$ ($d$ being the totally symmetric SU(N) color tensor). This adds up to 138 possible tensors for a general kinematical configuration. 

Thus, in order to simplify the calculation as much as possible without compromising the physics we want to describe, we choose the momentum configuration $(p_1,p_2,p_3,p_4)=(p,p,p,-3p)$. This particular choice has the following advantages:
\begin{enumerate}
	\item[\n{i}] It gives rise to loop integrals that are symmetric under the crossing of external legs, thus reducing the amount of diagrams one needs to evaluate;
	\item[\n{ii}] It allows to concentrate only on form factors multiplying tensor structures depending quadratically on the metric; all other possible structures will vanish when constructing the Landau gauge projectors $R_n^\s{T}$;
	\item[\n{iii}]  It is the only momentum configuration in which the 1-PR contributions to the connected four-gluon Green's function vanish, thus allowing to study the (projected) 1-PI component of the connected four-gluon vertex in isolation\footnote{This aspect would make the $(p,p,p,-3p)$ as the configuration of choice in a possible attempt to study this vertex on the lattice.}. 	
\end{enumerate}

Within this configuration, for a general SU(N) gauge group one has $3\times9=27$ possible tensor structures\footnote{The terms quadratic in the metric gives the 3 possible combinations $g_{\mu\nu} g_{\rho\sigma}$, $g_{\mu\rho} g_{\nu\sigma}$, and $g_{\mu\sigma} g_{\nu\rho}$; for the color structures on has 15 possibilities and 6 identities~\cite{Pascual:1980yu}, and therefore 9 independent tensors.}; however, for the special case of $N=3$, the additional identity~\cite{Pascual:1980yu}
\begin{equation}
\delta^{ab}\delta^{cd}+\delta^{ac}\delta^{bd}+\delta^{ad}\delta^{bc}=3[d_{abr}d_{cdr}+d_{acr}d_{bdr}+d_{adr}d_{bcr}],
\label{N3}
\end{equation}
further reduces the total number of tensorial combinations down to 24.  

When all this is combined with the one-loop dressed approximation employed, it turns out that the tensor structures to be considered are in fact only two, as the result can be cast in the form
\begin{equation}
	\left.\Gamma_{\mu\nu\rho\sigma}^{abcd}(p,p,p,-3p)\right\vert_{gg} =  V_{\Gamma^{(0)}}(p^2) \Gamma_{\mu\nu\rho\sigma}^{abcd(0)} + V_{G}(p^2) G^{abcd}_{\mu\nu\rho\sigma},
	\label{deco}
\end{equation}
where $\Gamma_{\mu\nu\rho\sigma}^{abcd(0)}$ is the tree-level tensor defined in~\noeq{4g}, while $G^{abcd}_{\mu\nu\rho\sigma}$ represents the totally symmetric tensor
\begin{equation}
	G^{abcd}_{\mu\nu\rho\sigma}=(\delta^{ab}\delta^{cd}+\delta^{ac}\delta^{bd}+\delta^{ad}\delta^{bc})(g_{\mu\nu}g_{\rho\sigma}+g_{\mu\rho}g_{\nu\sigma}+g_{\mu\sigma}g_{\nu\rho}).
	\label{Gtens}
\end{equation} 
In addition, the IR leading term coming from the ghost diagrams $(a)$ of~\fig{4g-1loop-dressed} only contributes to the latter structure, as one has
\begin{equation}
\sum_{i=1}^6\left.(a_i)^{abcd}_{\mu\nu\rho\sigma}\right\vert_{gg}=g^2G^{abcd}_{\mu\nu\rho\sigma} A(p^2),
\end{equation}
with
\begin{equation}
A(p^2)=-\frac92\frac1{d^2-1}\int_k\! k^2\left[1-\frac{(k\cdot\! p)^2}{k^2p^2}\right]^2
\frac{F(k)F(k+p)F(k+2p)F(k+3p)}{(k+p)^2(k+2p)^2(k+3p)^2}.
\label{Ap}
\end{equation}
Notice that as $p^2\to0$,~\1eq{Ap} yields
\begin{equation}
	A(p^2)\underset{p^2\to0}{\to}-\frac{9}{2d(d+2)}\int_k\!\frac{F^4(k^2)}{k^4},
\end{equation}
that is, the form factor $V_{G}(p^2)$ will develop a logarithmic IR divergence (in four-dimensions). The (rather lengthy) expressions for the remaining class of diagrams can be found in~\cite{Binosi:2014kka}.

\begin{figure}
\includegraphics[scale=0.975]{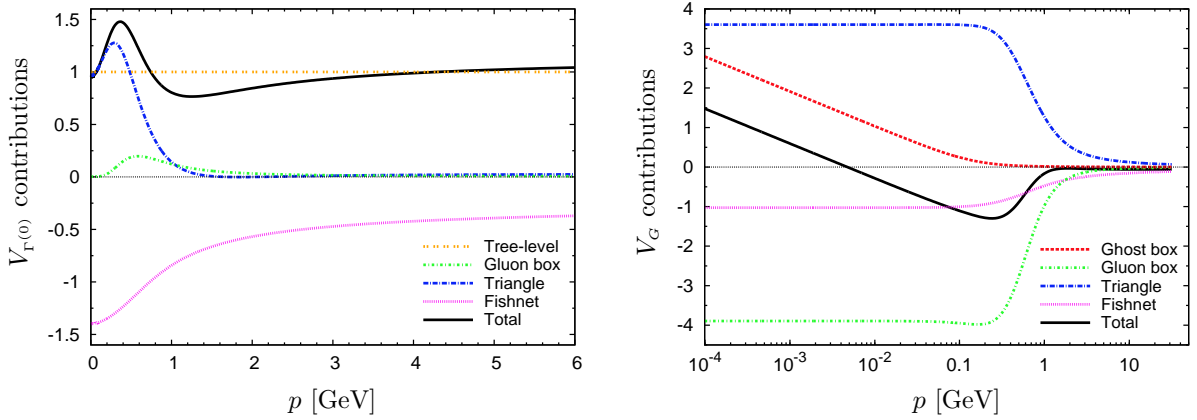}
\caption{\label{fig:G-Gamma0} (color online). Individual  one-loop dressed contributions to the tensor structure $\Gamma_{\mu\nu\rho\sigma}^{abcd(0)}$ (left panel) and $G^{abcd}_{\mu\nu\rho\sigma}$ (right panel). The black line coincides with the coefficient $V_{\Gamma^{(0)}}(p^2)$ and $V_{G}(p^2)$ respectively.}  
\end{figure}

The different contributions to the four-gluon vertex form factors can be evaluated using the functional fits to the quenched lattice data for the gluon and ghost two-point functions\footnote{In the case of the gluon propagator we have also employed a fit featuring the IR maximum discussed in Sect.~\ref{sect:2-point}; the results obtained were, however, independent from its presence.}. The results are shown in~\fig{fig:G-Gamma0}.

Consider first the left panel, where we show the contributions to $V_{\Gamma^{(0)}}$. These are purely gluonic in nature,  and all of them saturate in the IR. In particular we notice that, the contribution of the gluon boxes is negligible (indeed, as $p\to0$ it vanishes); the triangle terms feature a bump in the low momentum region, while the fishnet is negative. Adding everything up, one obtains the shape shown by the black line.

In the case of $V_G$ the situation is completely different (\fig{fig:G-Gamma0}, right panel). Gluon contributions are again saturating in the IR; however, in this case, the ghost boxes take over below few hundreds MeV$^2$, driving $V_G$ to an IR logarithmic divergence. As far as the remaining diagrams are concerned, gluon boxes are negative in this case; in addition, they are almost perfectly cancelled by the triangle contributions. When the negative contribution from the fishnet diagrams is finally added, one obtains the shape shown by the black line of~\fig{fig:G-Gamma0}; in particular, notice the presence of a zero crossing, a feature that was also present in the $R_3^{\Gamma^{(0)}}$ ratio previously studied in the case of the three-gluon vertex.

Now, the result~\noeq{deco} is due to the one-loop dressed approximation employed; a general analysis based on Bose symmetry~\cite{Binosi:2014kka}, shows that the terms quadratic in the metric contributing to the four-gluon vertex in the $(p,p,p,-3p)$ momentum configuration allow for an extra tensor, namely one has
\begin{equation}
	\left.\Gamma_{\mu\nu\rho\sigma}^{abcd}(p,p,p,-3p)\right\vert_{gg} =  V_{\Gamma^{(0)}}(p^2) \Gamma_{\mu\nu\rho\sigma}^{abcd(0)} + V_{G}(p^2) G^{abcd}_{\mu\nu\rho\sigma}+V_{X'}(p^2)X^{'abcd}_{\mu\nu\rho\sigma},
	\label{deco1}
\end{equation}
where\footnote{Bose symmetry alone does not permit to fix completely the tensor $X^{'abcd}_{\mu\nu\rho\sigma}$; its explicit form~\noeq{Xprime} is obtained by requiring that the latter should be orthogonal to the tensor $G$, that is $$G^{abcd}_{\alpha\beta\rho\gamma} P^{\mu\alpha}(p)P^{\nu\beta}(p)P^{\rho\gamma}(p)P^{\sigma\delta}(p)X^{'abcd}_{\alpha\beta\rho\gamma}=0.$$}
\begin{eqnarray}
	X^{'abcd}_{\mu\nu\rho\sigma}&=&g_{\mu\nu}g_{\rho\sigma}(\delta^{ab}\delta^{cd}/3-d^{abr}d^{cdr})+ g_{\mu\rho}g_{\nu\sigma}(\delta^{ac}\delta^{bd}/3 -d^{acr}d^{bdr})\nonumber \\
	&+& g_{\mu\sigma}g_{\nu\rho}(\delta^{ad}\delta^{bc}/3-d^{adr}d^{bcr}).
	\label{Xprime}
\end{eqnarray}

This means that a possible lattice evaluation of the connected four-gluon function in this particular momentum configuration completely determines the terms of the four-gluon vertex quadratic in the metric tensor, through the  measurement of the ratios
\begin{equation}
	R_4^{\Gamma^{(0)}}(p^2)=V_{\Gamma^{(0)}}(p^2)+\frac1{81}V_{X'}(p^2);\qquad R_4^G(p^2)=V_G(p^2);\qquad R_4^{X'}(p^2)=V_{X'}(p^2)+\frac9{164}V_{\Gamma^{(0)}}(p^2).
\end{equation}
According to our description one expects $V_{\Gamma^{(0)}}(p^2)$ to be finite and $V_G(p^2)$ to display an IR divergence; nothing can be however said at the moment for the form factor $V_{X'}$; however the vanishing of this latter quantity in the one-loop dressed approximation, points towards its finiteness. Thus,  we would expect the measurement of only one divergent ratio, and namely $R_4^G(p^2)$. 

\section{Conclusions}

The gluon and ghost field display a very different behavior in the deep IR: the latter remains nonperturbatively massless, whereas the former acquires a dynamically generated mass. This fact, which has been unequivocally established in the Landau gauge using discrete as well as continuous methods, turns out to have a profound impact on the $n$-gluon sector of the theory, as diagrams involving ghost loops gives unavoidably origin to IR divergences.   

In the case of the 2-gluon sector it is the gluon inverse dressing function $J(q^2)$  that shows such a divergence (with $J(q^2)\sim\log q^2$ in four dimensions); while the presence of such a divergence does not interfere with the finiteness of the gluon 2-point function (for the gluon full kinetic term is multiplied by $q^2$), it nevertheless implies that the full propagator has an IR maximum located at $q=q_\s{\Delta}$, and, correspondingly, $q^2J(q^2)$ has a minimum, located at $q=q_\s{J}$.

For the three-gluon sector a (negative) IR divergence appears when projecting the full three-gluon vertex onto its tree level value, and choose the so-called orthogonal momentum configuration. Due to the relation~\noeq{R-J}, the location of the point $q=q_0$ at which $R_3^{\Gamma^{(0)}}(q^2,0,\pi/2)$ crosses zero and turns negative can be roughly estimated from $q_\s{J}$ and turns out to be around $130$ MeV ($45$ MeV)  for the $N=3$ ($N=2$) case. 

An IR divergence appears also when evaluating the four-gluon vertex in the $(p,p,p,-3p)$ momentum configuration, even though in this case it does not manifest in the projector $R_4^{\Gamma^{(0)}}(p^2)$ onto the tree-level tensor, rather in $R_4^{G}(p^2)$, where $G$ is the totally symmetric tensor~\noeq{Gtens}. 

The picture presented here, elaborated within the PT-BFM formalism which allows for a gauge-invariant separation of ghost and gluon contributions to the gluon propagator, is found to be in agreement with lattice data whenever the latter are available. 

Two are the questions that needs to be addressed. 

To begin with, since Green's functions depend on the gauge fixing employed, it would be important to evaluate them in different gauges in order to ascertain what aspects of their nonperturbative behavior are affected by a change of gauge. This is particularly relevant in the 2-point sector, as a recent preliminary study using a combination of SDE and Nielsen identities\footnote{These identities express the gauge-dependence of ordinary Green’s functions (propagators, vertices, etc.) in terms of special auxiliary functions associated with an extended BRST sector~\cite{Nielsen:1975fs,Nielsen:1975ph}.} has revealed that in the renormalizable $\xi$ gauges the ghost dressing function vanishes in the deep IR~\cite{Aguilar:2015nqa}. If this result persists refined studies, possibly including lattice simulations, its impact on the IR behavior of ancestor ghost loops (and consequently the $n$-gluon sector of the theory) needs to be thoroughly addressed.    

Second, one would like to see wether or not the presence of zero crossings and IR divergences has some impact on hadron phenomenology ({\it e.g.}, the hadron spectrum), in order to connect results on the theory's most basic building blocks with its observables properties, along the lines recently discussed in~\cite{Binosi:2014aea}. At a first sight it would look like the zero crossing momentum $q_0$ is located too deep in the IR for both the three- and four-gluon vertex to affect the Bethe-Salpeter equations one needs to solve. However, a preliminary analysis of unquenching effects shows that dynamical quarks have the tendency to move $q_0$ closer to the phenomenologically relevant region of few hundreds MeV. This is relatively easy to understand, as the main effect of adding dynamical quarks is to suppress the saturation point of the gluon propagator while leaving the ghost dressing function practically invariant~\cite{Ayala:2012pb,Aguilar:2012rz,Aguilar:2013hoa};  consequently gluonic contributions will be suppressed whereas ghost contributions will be of the same size of the unquenched ones, which results in pushing the zero crossing towards higher momentum values with respect to the quenched case. 

\bigskip

\section*{Acknowledgements}

I would like to thank the organizers of DISCRETE 2014 conference for their kind invitation and hospitality.

\bigskip


\providecommand{\newblock}{}

\end{document}